\documentclass[twocolumn,showpacs,preprintnumbers,amsmath,amssymb,floatfix]{revtex4}
\input{epsf}

\usepackage{graphicx}
\usepackage{dcolumn}
\usepackage{bm}

\begin{document}

\preprint{APS/123-QED}

\title{Precision measurement with an optical Josephson junction}

\author{H. T. Ng$^{1}$, K. Burnett$^{1}$ and J. A. Dunningham$^{2}$}
\affiliation{$^{1}${Clarendon Laboratory, Department of Physics,
University of
Oxford,}\\
{Parks Road, Oxford OX1 3PU, United Kingdom}\\ $^{2}${School of
Physics and Astronomy, University of Leeds, LS2 9LT, United
Kingdom}}
\date{\today}

\begin{abstract}
We study a new type of Josephson device, the so-called ``optical
Josephson junction'' as proposed in Phys. Rev. Lett. {\bf 95},
170402 (2005).  Two condensates are optically coupled through a
waveguide by a pair of Bragg beams. This optical Josephson junction
is analogous to the usual Josephson junction of two condensates
weakly coupled via tunneling.  We discuss the use of this optical
Josephson junction, for making precision measurements.
\end{abstract}

\pacs{03.75.Lm, 03.75.Gg, 03.75.-b} \maketitle

\section{Introduction}
Atom optics has obtained an unprecedented development since the
advent of Bose-Einstein condensates (BEC's) of dilute alkali atomic
gases \cite{Bradley}. For instance, the applications of atom optics
have been realized such as atom laser, solition generation
\cite{Rolston} and matter wave interferometry \cite{Shin0}. Besides,
many new techniques of manipulating the cold atoms have been devised
such as Bragg scattering \cite{Kozuma} and magnetic waveguide
\cite{Leanhardt}, etc. These sophisticated techniques may lead to
reaching the goal in higher precision measurement \cite{Dunningham}
and applications in quantum information processing \cite{Nielsen}.

Recently, Saba {\it et. al} \cite{Saba} and Shin {\it et. al}
\cite{Shin} have demonstrated that the Josephson effect by using the
technique of atom optics, Bragg scattering.  Two beams of atoms are
optically extracted from two separate trapped Bose-Einstein
condensates (BEC's) using Bragg scattering \cite{Kozuma}. Then,
these two beams of atoms overlap and interfere with each other and
the measurement process creates a relative phase between the two
BEC's. They showed that the Josephson coupling of these two
spatially separate systems can be made through an intermediate
coupling system. It is quite different to the conventional Josephson
devices, such as superconducting systems \cite{Anderson}, and
Bose-Einstein condensates \cite{Cataliotti}, in which the two
quantum systems are spatially connected to each other and the
wavefunction of them have small overlap.

In this paper, we investigate that the two trapped condensates are
connected by outcoupling a small fraction of condensates via the
Bragg scattering \cite{Kozuma} and those outcoupling atoms are
transported through a magnetic waveguide \cite{Leanhardt} to
replenish the trapped condensates \cite{Shin}. We show that this
system is equivalent to the usual Hamiltonian describing the
Josephson effect of BEC's trapped in a double-well potential. The
strength of the Josephson coupling can be explicitly controlled by
varying the outcoupling rate of atoms using the Bragg beams whereas
the coupling phase between these two condensates can be tuned by
adjusting the phase shifts of the outcoupled atoms \cite{Shin}.
Moreover, the spontaneous scattering due to the Bragg beams is
negligible by virtue of the large detuning of the pulses. This
``optical Josephson junction'' could be used, for example, to
implement the precision scheme proposed by Dunningham and Burnett
\cite{Dunningham} in which they consider the precision measurement
using an entangled BEC's trapped in a double-well potential.

We now study the implementation of the precision measurement scheme
being suggested by Dunningham {\it et al.} \cite{Dunningham}, which
can measure the nonlinear interaction strength and gravity, using
this ``optical Josephson junction''. According to this proposal, the
measurement of phase can be done by appropriately turning on and off
the Josephson coupling. Then, the final phase information is encoded
in the number fluctuation that can be detected from the collapse and
revival of the relative phase between the condensates. This scheme
provides a simple way to measure phase with Heisenberg limited
accuracy, i.e., the phase uncertainty scales as $1/N$, where $N$ is
the number of atoms.

\begin{figure}[ht]
\includegraphics[height=6cm]{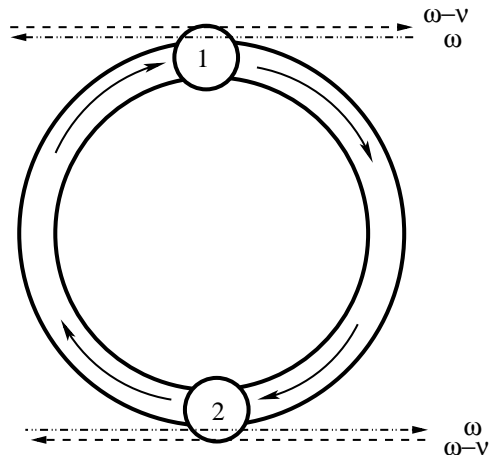}
\caption{\label{fig1} The schematic of the optical  coupling of two
trapped condensates 1 and 2, and the outcoupling atoms are
transferred via a waveguide in the ring form. The pair of Bragg
beams with frequencies $\omega$ and $\omega-\nu$ are denoted by
dash-dot line and dash line respectively.}
\end{figure}

The paper is organized as follows: In Sec. II, we introduce the
model of this ``optical Josephson junction''. In Sec. III, we derive
an effective Hamiltonian under the two-mode approximation of two
trapped BEC's. In Sec. IV, we study the implementation of Dunningham
{\it et al.} scheme in this system.  We briefly review the mechanism
of the Bragg scattering in appendix A.

\section{Basic model}
We consider two trapped condensates connected to a 1D ring-form
waveguide using the $M$-th order Bragg scattering \cite{Saba,Shin}
as shown in Fig 1.   In fact, this Bragg scattering is a $2M$-th
order multi-photon Raman process \cite{Kozuma,Berman}. The
stimulated emission and stimulated absorption are made by using the
two Bragg beams being acted as a pump and probe field with
counter-propagation and slightly frequency difference, i.e.,
$\omega=$ and $\omega-\nu$.  The wavelength and wave number of the
Bragg beam with frequency $\omega$ are $\lambda$ and
$k=2\pi/\lambda$ respectively.

This pair of Bragg beams give out a certain amount of momentum to
the two trapped BEC's $1$ and $2$ respectively. Then, a small amount
of BEC's are outcoupled from these two trapped BEC's with definite
momenta ${\hbar}k_1$ and ${\hbar}k_2$ at the $M$-th order Raman
process. Without loss of generality, we assume the two momenta $k_1$
and $k_2$ with the equal magnitude but opposite sign, for
$\hbar^2k^2_{1,2}/2m=M\hbar\nu$ and $m$ is the mass of an atom.
These outcoupling condensates are then transported via the waveguide
to the other trapped BEC. This conservation of total number of
particles may enable us to overcome the limitation of the coupling
time in the experiments \cite{Saba,Shin} in which the atoms were
linearly coupled out and not transferred back to the system
\cite{Shin}. But now the coupling process can be continued if the
number of atoms are conserved in the system.

We discuss the Hamiltonian of this system and derive an effective
two-state Hamiltonian to describe the Josephson effect. The
Hamiltonian of the total system has the form
\begin{eqnarray}
H&=&H_0+H_{\rm ring}+H_{\rm couple},
\end{eqnarray}
where $H_0$, $H_{\rm ring}$ and $H_{\rm couple}$  are the
Hamiltonians of; BEC's 1 and 2, the outcoupled atoms in the 1D ring,
and the coupling between the trapped atoms and the ring
respectively. The Hamiltonian for the trapped atoms has the form and
can be expressed in terms of the arc length, $s=r\phi$ for $r$ is
the radius of the ring and $\phi$ is the polar angle of the ring,
\begin{eqnarray}
H_0&=&\sum^2_{j=1}\oint{ds}\Bigg\{\Psi^{\dag}_j(s)\Bigg[-\frac{\hbar^2}{2m}\frac{\partial^2}{\partial{s^2}}+V_j(s)\Bigg]
\Psi_j(s)\nonumber\\
&&+\frac{U_0}{2}\Psi^\dag_j(s)\Psi^\dag_j(s)\Psi_j(s)\Psi_j(s)\Bigg\},
\end{eqnarray}
where $\Psi_{j}(s)$ and $V_j(s)$ are the field operator and trapping
potential respectively for condensate $j$, $U_0$ is the interaction
strength and $j=1,2$.

We consider that the two condensates are held in two different traps
that are sufficiently deep, and hence the single-mode approximation
can be applied \cite{Milburn}. This allows us to describe the
trapped condensates as the local ground state of the trap. The field
operator $\Psi_j(s)$ can then be approximated by ${c_j}\psi_j(s)$,
where $c_j$ and $\psi_j(s)$ are the annihilation operators and the
mode function of the potential of the $j$-th trap and $j=1,2$. If we
take the system to be symmetric so that $\Psi_1(s)$ has the same
form as $\Psi_2(s)$, the Hamiltonian $H_0$ can be written as
\begin{eqnarray}
H_0&=&E_0(c^\dag_1c_1+c^\dag_2c_2)+\frac{\hbar\kappa}{2}[(c^\dag_1c_1)^2+(c^\dag_2c_2)^2],
\end{eqnarray}
where
\begin{eqnarray}
E_0&=&{\oint}ds\psi^*_j(s)\Bigg[-\frac{\hbar^2}{2m}\frac{\partial^2}{\partial{s^2}}+V_j(s)\Bigg]\psi_j(s),
\end{eqnarray}
is the eigenenergy of the two modes and
\begin{eqnarray}
\kappa&=&U_0\oint{ds}|\psi^*_j(s)\psi_j(s)|^2,
\end{eqnarray}
is the self-interaction strength.

The free condensates can be outcoupled from the trapped condensate
using the Bragg scattering \cite{Kozuma}. We briefly review its
mechanism in the appendix A.  The nearly perfect efficiency of the
first order Bragg resonance has been observed experimentally
\cite{Kozuma}. However, the efficiency of the high order Bragg
scattering is decreased significantly \cite{Kozuma}. The Hamiltonian
representing the effective coupling between the trapped atoms and
the free condensate is written, thus,
\begin{equation}
H_{\rm couple}=\gamma\oint{ds}[\Psi^\dag_f(s)\Psi_1(s)
+{e^{-i\pi\delta{k}r}}\Psi^\dag_f(s)\Psi_2(s)]+{\rm H.c.},
\end{equation}
where ${\gamma}$ is the outcoupling amplitude between two trapped
BEC's and free outcoupling condensates \cite{Choi},
$\delta{k}=k_1+k_2-4k$ \cite{Shin}respectively.  Here, we can
consider the coupling between the trapped condensates and the free
condensates with momenta $k_{1,2}$.  Moreover, we consider the
replenishment process in which the free condensates are
deaccelerated by the Bragg beams and merge into the trapped
condensates \cite{Chikkatur}. We assume that the excitation in this
replenishing process is negligible.  On the other hand, the relative
phase shift $\pi\delta{k}r$ is generated during the flight between
the two condensates.  The phase shift $\delta{k}$ directly depends
on the Bragg beams and the momenta $k_{1,2}$ which can be adjusted
in the experiment \cite{Shin}.

The Hamiltonian of the 1D ring has the corresponding form:
\begin{eqnarray}
H_{\rm
ring}&=&\oint{ds}\Bigg[\Psi^{\dag}_f(s)\Bigg(-\frac{\hbar^2}{2m}\frac{\partial^2}{\partial{s^2}}\Bigg)
\Psi_f(s)\nonumber\\
&&+\frac{U_0}{2}\Psi^\dag_f(s)\Psi^\dag_f(s)\Psi_f(s)\Psi_f(s)\Bigg],
\end{eqnarray}
where $\Psi_f(s)$ is the field operator of the condensate in this
ring-form waveguide transported by a waveguide. We approximate these
unconfined in the angular direction, but confined radially and
weakly interacting condensate as the freely propagating
non-interacting condensates.  We can justify this approximation by
considering the magnitude of the kinetic energy and the strength of
the nonlinear interaction. The kinetic energy $\hbar\omega_{k_j}$ of
the outcoupled condensates by the Bragg beams is with
$\omega_{k_j}{\sim}10^2$ kHz \cite{Shin} whereas the nonlinear
interaction of this nearly free condensates $n_0U_0/\hbar$ is about
${1}$  kHz \cite{Leggett}.  Here $n_0$ is density of the outcoupled
condensate which must satisfy $n_0a_s{\ll}1$ and $a_s$ is the
scattering length of the atom.    From the above estimation, we can
argue that this weakly interacting outcoupled condensates as free
condensates.

The condensates in the ring are composed of two free condensates
outcoupled from the two trapped condensates as
\begin{eqnarray}
\Psi_f(s)&=&{(2\pi{r})}^{-1/2}(e^{ik_1s}g_{k_1}+e^{ik_2s}g_{k_2}),
\end{eqnarray}
where $2\pi{r}$ is the circumference of the loop, $k_{1,2}$ should
satisfy the boundary condition $k_{1,2}r={\alpha}\hbar$ and $\alpha$
is an integer. Then, the Hamiltonian of the ring can be written as
\begin{eqnarray}
H_{\rm ring}&=&\sum^2_{j=1}\hbar\omega_{k_j}g^\dag_{k_j}g_{k_j},
\end{eqnarray}
where $\omega_{k_i}={\hbar}k^2_i/2m$.

\section{Effective Hamiltonian}
We now derive an effective Hamiltonian to describe the Josephson
effect between two trapped BEC's.  The free propagating condensates
act as immediate states to couple the two trapped condensates.  The
coupling Hamiltonian is given by
\begin{equation}
H_{\rm
couple}=\hbar\sum^2_{j=1}\gamma'(k_j)(c^\dag_1+e^{i\theta}c^\dag_2)g_{k_j}+{\rm
H.c.},
\end{equation}
where
$\gamma'(k_j)=\gamma(2\pi{r})^{-1/2}\oint{ds}e^{-i{k_j}s}\psi_{j}(s)$,
$j=1,2$ and $\theta=\pi\delta{k}r$. We assume that the two coupling
strengths $\gamma'(k_j)$ are similar by letting
$\oint{ds}e^{-ik_1s}\psi_{1}(s)\approx\oint{ds}e^{-ik_2s}\psi_{2}(s)$.
The Heisenberg equation of motion of $g_{k_j}$ is
\begin{eqnarray}
i\dot{g}_{k_j}&=&\omega_{k_j}g_{k_j}+\gamma'(k_j)(c_1+e^{-i\theta}c_2).
\end{eqnarray}
We use the adiabatic approximation, i.e., $\dot{g}_{k_j}{\approx}0$
as the condition $\omega_{k_j}{\gg}|\gamma'(k_j)|$ is satisfied:
\begin{eqnarray}
\label{outstate}
g_{k_j}&\approx&-\frac{\gamma'(k_j)}{\omega_{k_j}}(c_1+e^{-i\theta}c_2).
\end{eqnarray}
We can see that it is true if we can assure that $\gamma$ is much
less than $\omega_{k_{1,2}}$. Physically speaking, the validity of
this adiabatic approximation assumes that the transition time
between these intermediate states and the two trapped states is
short enough.  Moreover, it is noted that the whole $2M$-th order
Raman process requires a finite time duration with the interaction
of laser to be completed.  This time duration must be short compared
to the time of the Josephson coupling.

The resulting effective two-state Hamiltonian can thus be written as
\begin{eqnarray}
H_{\rm
eff}&=&\Bigg(E_0-\frac{{\hbar}g}{2}\Bigg)(c^\dag_1c_1+c^\dag_2c_2)+\frac{\hbar\kappa}{2}[(c^\dag_1c_1)^2+(c^\dag_2c_2)^2]\nonumber\\
&&-\frac{{\hbar}g}{2}\Big(e^{-i\theta}c^\dag_1c_2+e^{i\theta}c^\dag_2c_1\Big),
\end{eqnarray}
where $g=2\sum^2_{j=1}{\gamma'}^2(k_j)/\omega_{k_j}$. It is
noteworthy that this effective two-mode Hamiltonian is akin to the
Josephson Hamiltonian of the external \cite{Milburn} and the
internal \cite{Sorensen} BEC systems.  This optical Josephson
coupling can then yield the Josephson effect of two condensates.
Likewise, this Josephson coupling can be controlled by the strength
of the Bragg beam to vary $g$ whereas the phase of Josephson
coupling can be adjusted by the phase shift $\theta$ that can be
done by changing $k$ and the arc length between the BEC's. Indeed,
this feature is very useful to the application of this Josephson
junction.

It is convenient to write the effective Hamiltonian in terms of the
angular momentum operators as
\begin{eqnarray}
J_x&=&\frac{1}{2}(c^\dag_1c_2+c^\dag_2c_1),\\
J_y&=&\frac{1}{2i}(c^\dag_1c_2-c^\dag_2c_1),\\
J_z&=&\frac{1}{2}(c^\dag_2c_2-c^\dag_1c_1),
\end{eqnarray}
where the total atom numbers $N=c^\dag_2c_2+c^\dag_1c_1$ is
conserved. Then, the Hamiltonian is written as
\begin{eqnarray}
H_{\rm
eff}&=&\hbar\kappa{J^2_z}+{\hbar}g(\cos\theta{J_x}+\sin\theta{J_y})+C,
\end{eqnarray}
where $C=(E_0-{\hbar}g/2)N+\kappa{N^2}/4$. This constant $C$ does
not affect the quantum dynamics of the system.  Therefore, we will
ignore it in the subsequent discussion.

\section{Implementation of precision measurement}
In this section, we discuss the implementation of precision
measurement on the optical Josephson junction.  Dunningham and
Burnett proposed a precision measurement scheme using a
number-squeezed BEC trapped in a double-well potential
\cite{Dunningham}. This scheme requires the active control of the
Josephson coupling.  The phase information can be finally obtained
by measuring the number fluctuations which can be detected via the
visibility of the interference fringes of the two condensates.
However, there are some limitations of the BEC double-well system
with respect to the dynamical control of the Josephson coupling. The
Josephson coupling strength exponentially depends on the height of
potential barrier between the two wells. The potential barrier has
to be tuned very accurately in order to vary the coupling.

Besides, when the Josephson coupling is large, there needs to be a
large spatial overlap between the two localized wave functions in
two wells.  In this case, the usual two-mode approximation is no
longer valid. This problem can now be avoided in the optical
Josephson junction system due to the large separation between the
two trapped condensates. Since the optical Josephson junction is a
well-controlled system, it is a promising model to implement this
scheme.

We proceed to describe how to realize the protocol of the scheme
proposed by Dunningham {\it et al.}. If we choose the phase shift
$\theta=\pi$, the effective Hamiltonian has the form,
\begin{eqnarray}
H_{\rm eff}&=&\hbar\kappa{J^2_z}+{\hbar}g{J_x}.
\end{eqnarray}
We first prepare the initial state as a number-squeezed state
$|N/2\rangle_1|N/2\rangle_2$ which contains a definite number of
atoms trapped in two wells respectively.  This can be prepared by
adiabatically switching off the Josephson coupling strength so that
the condensates are isolated in two different traps in the Fock
regime, i.e., $g\ll{\kappa{N}}$ \cite{Leggett}.

We can use this system to measure a relative phase between the
condensates as follows.  A large Josephson coupling is turned on
rapidly with the strength $g\gg{\kappa{N}}$ by using Bragg pulses
with $g=g^*$. It is convenient to write the state in the new
eigenbasis, i.e., the symmetric mode $\alpha=(c_1+c_2)/\sqrt{2}$ and
antisymmetric mode $\beta=(c_1-c_2)/\sqrt{2}$ with respect to two
traps. The quantum state is in this new basis, has the form
\cite{Dunningham}
\begin{eqnarray}
|\Psi\rangle&=&\sum^{N/2}_{m=0}(-1)^{m}C_m|2m\rangle_\alpha|N-2m\rangle_\beta,
\end{eqnarray}
where $C_m=\sqrt{(2m)!(N-2m)!}/2^{N/2}m!(N/2-m)!$. It is worth
noting that this superposition of states is robust against the
particle loss \cite{Dunningham2}.  In this regime, there is a small
energy difference, $g^*$, between the symmetric and antisymmetric
modes, different phases result for the terms in the superposition.
Then, the system is held for a certain time $t=t^*$ to allow the
natural evolution of the system.

Next, the Josephson coupling is suddenly switched off fast enough
with respect to coupling between wells, but slow with respect to
energy level spacing in each well.  Thus, the state is conveniently
expressed in terms of the number basis in each well. The quantum
state becomes \cite{Dunningham}
\begin{eqnarray}
\label{state_phi}
|\Psi\rangle&=&\frac{1}{2^{N/2}{(N/2)}!}\sum^{N}_{n=0}(-1)^nD_n|n\rangle_1|N-n\rangle_2,
\end{eqnarray}
where
\begin{widetext}
\begin{eqnarray}
D_n=\sum^{k/2}_{p={\rm max}\{0,n-{N}/{2}\}}\left(
\begin{array}{c} {N}/{2}
\\ p^* \end{array} \right)  \left( \begin{array}{c} p^* \\
p \end{array} \right)
\sqrt{n!(N-n)!}(i\sin\phi)^{p^*}(2\cos\phi)^{n-2p}
\end{eqnarray}
\end{widetext}
for $p^*=N/2-n+2p$ and $\phi={\hbar}g^*{t^*}$. This completes the
measurement of the phase $\phi$ which is recorded in the quantum
state now. The information of this phase $\phi$ can be obtained from
the number uncertainty ${\Delta}n$, where
$n=c^\dag_1c_1=N-c^\dag_2c_2$ is the number of atoms in the trap 1
(or 2).  From Eq. (\ref{state_phi}), the number uncertainty can be
calculated as \cite{Dunningham}
\begin{eqnarray}
\Delta{n}&=&\frac{N}{2\sqrt{2}}\sin\phi.
\end{eqnarray}
This number uncertainty is of order $N$. According to the
number-phase uncertainty relation $\Delta{n}\Delta{\phi}\sim{1}$, we
can see that the uncertainty of the phase $\phi$, for the minimum
uncertainty state, is of order $N^{-1}$, i.e., Heisenberg limited.

This number variance can be experimentally determined from the
interference pattern. Bragg scattering provides a non-destructive
method to determine the relative phase. A small fraction of
condensates are coupled out horizontally from these two condensates
and allowed to interfere with each other.  The relative phase
between the two trapped condensates can be determined from the
interference patten of these two outcoupled condensates
\cite{Wright,Dunningham3}. The interference pattern of the fringes
between the two outcoupled condensates \cite{Sinatra},
$I=\int{dx}[\Psi^\dag_{f1}(x)\Psi_{f2}(x)+\Psi^\dag_{f2}(x)\Psi_{f1}(x)]$,
is directly proportional to the interference terms of two trapped
condensates, where $\Psi_{f1}$ and $\Psi_{f2}$ are the field
operator of two outcoupled condensates from the $j$-th trap, and $x$
is coordinate in the horizontal direction. Following the similar
treatment to the previous section and taking $\theta=\pi$,  the
intensity of the interference fringes, $I$, of these two outcoupled
condensates can be obtained as \cite{Sinatra}
\begin{eqnarray}
I=\frac{\gamma'(k_1)\gamma'(k_2)}{\omega_{k1}\omega_{k_2}}\langle\psi(\tau)|{c}^\dag_1c_2+{c}^\dag_2c_1|\psi(\tau)\rangle,
\end{eqnarray}
where $\tau$ is the time of holding the system with the nonlinear
interaction of the atoms.  This state vector is given by
\begin{widetext}
\begin{equation}
|\psi(\tau)\rangle=\frac{1}{2^{N/2}{(N/2)}!}\sum^{N}_{n=0}(-1)^ne^{-i{\hbar\kappa}[n^2+(N-n)^2]\tau/2}D_n|n\rangle_1|N-n\rangle_2.
\end{equation}
\end{widetext}
Thus, the intensity, $I$, is proportional to
\begin{eqnarray}
I(\tau)&\propto&\sum_{n}D^*_{n+1}D_n\sqrt{(n+1)(N-n)}e^{i\hbar\kappa(2n+1-N){\tau}}\nonumber\\
&&+D^*_{n-1}D_n\sqrt{n(N-n+1)}e^{i\hbar\kappa(N-2n+1){\tau}}.
\end{eqnarray}
The collapse times $t_{\rm coll}$ can be estimated by considering
the particle numbers in the range, $n=N/2\pm\Delta{n}/2$. Hence, the
collapse times of the relative phase $t_{\rm coll}$ is about
$\pi/2\hbar\kappa\Delta{n}$ \cite{Wright,Dunningham3}. This collapse
time can be determined by holding the system with the nonlinear
interaction as a function of time $\tau$ and measure the
corresponding intensity $I$ with different $\tau$'s. Therefore, we
can determine the number fluctuation and hence the required phase
information.

From an experimental point of view, the collapse time is very short
and so the damping and decoherence effects may be minimized. Hence,
the number variance may be able to be measured accurately by this
method.   Although the revival time can reveal the phase
information, it takes a much longer time to observe.  Bear in mind
that the measurement of collapse and revival time of a Bose-Einstein
condensate has been demonstrated in the experiment \cite{Greiner}.

\section{Discussion}
It is noted that the low efficiency of the higher order Bragg
scattering may limit the effective coupling of the two coupled
BEC's. Nevertheless, we can still justify the validity of this
optical based Josephson junction as discussed above. On the other
hand, it is very interesting to compare this optical based Josephson
junction to the solid-state counterpart.

In summary, we theoretically study the microscopic model of  the
``optical Josephson junction'' and derive the effective Hamiltonian
of this device. We have discussed how this system can be used to
implement Hesienberg limited precision measurement. This measurement
can be made by detecting the collapse time non-destructively.

\acknowledgments H.T.N. was supported by the Croucher Foundation.
K.B. thanks the Royal Society and Wolfson Foundation for support.

\appendix
\section{Bragg scattering}
In this appendix, the basic mechanism of the condensates using the
Bragg scattering is briefly reviewed. The pump-probe mechanism has
been discussed in detail in the reference \cite{Berman}.  The pump
and probe fields impart a momentum $2k$ to the ground state of the
condensates each time by coupling to the excited state with a large
detuning $\Delta=\tilde{\omega}-\omega$ between the two-level atoms
with the energy difference $\hbar\tilde{\omega}$ and the laser
field. To elucidate this process, we study the first order Bragg
resonance case by considering a time-independent Hamiltonian in the
interaction picture and assume the zero ground state energy for
$g_0$ which is given by
\begin{eqnarray}
H_{\rm 1
st}&=&\hbar{\Delta}_1{e}^\dag_ke_k+\hbar({\Delta}_1-{\Delta}_2)g^\dag_{2k}g_{2k}+\hbar\Omega(g^\dag_0e_k+e^\dag_kg_0)\nonumber\\
&&+\hbar\Omega'(g^\dag_{2k}e_k+e_kg^\dag_{2k}),
\end{eqnarray}
where $e_{nk}$ and $g_{nk}$ are the annihilation operators for the
excited and ground states with the momentum $nk$, $\Omega$ and
$\Omega'$ are the coupling strength of the pump field and the probe
field of this pair of Bragg beams; and
${{\Delta}}_1=\Delta+\omega_k$ and
${\Delta}_2=\Delta+\omega_{k}-\omega_{2k}+\nu$ are the detuning
between the ground and the excited states with the different momenta
and the Bragg beams, for $\omega_{nk}=\hbar(nk)^2/2m$.

The equations of motion for the different momentum states are given
by:
\begin{eqnarray}
i\dot{g}_0&=&\Omega{e}_k, \\
i\dot{e}_k&=&{\Delta}_1e_k+\Omega{g}_0+\Omega'g_{2k},\\
i\dot{g}_{2k}&=&({\Delta}_1-{\Delta}_2)g_{2k}+\Omega'e_{2k}.
\end{eqnarray}
At the Bragg resonance, the detuning
${\Delta}_1-{\Delta}_2=4\omega_{k}-\nu$ equals zero at
$\nu=4\omega_k$. The excited state $e_{k}$ can be adiabatically
eliminated as ${\Delta}_1\gg\Omega$,$\Omega'$, i.e.,
\begin{eqnarray}
e_{k}&=&-\frac{1}{{\Delta}_1}(\Omega{g}_0+\Omega'g_{2k}).
\end{eqnarray}
Therefore, the equations of motion for these two ground states with
different moment have the form:
\begin{eqnarray}
i\dot{g}_{0}&=&-\frac{1}{{\Delta}_1}(\Omega^2g_0+\Omega\Omega'g_{2k}),\\
i\dot{g}_{2k}&=&-\frac{1}{{\Delta}_1}(\Omega\Omega'g_0+\Omega'^2g_{2k}).
\end{eqnarray}
Clearly, we can see that these two momentum states are effectively
coupled with each other at the first order Bragg resonance.

In general, we can consider the $M$-th order Bragg scattering which
is a $2M$-th multi-photon Raman process.  Within this process, the
different momentum modes are virtually excited but they can be
adiabatically eliminated because of energy conservation being
unfavourable. It is legitimate to consider the effective coupling
between the trapped condensates and the free momentum states
$\hbar{\omega_{k_{1,2}}}=M{\nu}$ at the Bragg resonance only in
which the energy is conserved.  The explicit form of the effective
coupling between the trapped and free states, ${\gamma}$, has been
found as \cite{Berman}:
\begin{eqnarray}
{\gamma}&=&\Bigg|\frac{(\Omega\Omega'/{\Delta})^M}{[(M-1)!]^2\omega^{M-1}_{2k}}
\Bigg|.
\end{eqnarray}
The detailed analysis can be referred to the reference
\cite{Berman}.

\newpage

\end{document}